\documentstyle[10pt,psfig]{article}
\begin{document}
\begin{center} 
{\bf Speckle interferometry}\\
\end{center}
\vspace{1cm}

\begin{center} 
{\bf Swapan K Saha}\\
Indian Institute of Astrophysics \\
Bangalore 560 034\\
India\\
e-mail: sks@iiap.ernet.in\\
\end{center}
\vspace{1cm}

\noindent
{\bf Abstract} :
We have presented the basic mathematical treatment of interferometry in the
optical domain. Its applications in astronomical observations using both the
single aperture, as well as the diluted apertures are described in detail. 
We have also described about the shortcomings of this technique in the presence 
of Earth's atmosphere. A short descriptions of the atmospheric turbulence
and its effect on the flat wavefront from a stellar source is given. The 
formation of speckle which acts as carrier of information is defined. 
Laboratory experiments with phase modulation screens, as well as the resultant
intensity distributions due to point source are demonstrated. The experimental 
method to freeze the speckles, as well as data processing techniques for both 
Fourier modulus and Fourier phase are described. We have also discussed the 
technique of the aperture synthesis using non-redundant aperture masks at the
pupil plane of the telescope, emphasizing set on the comparison
with speckle interferometry. The various methods of image restoration and 
their comparisons are also discussed. Finally, 
we have touched upon certain astrophysical problems which can be tackled with 
the newly developed speckle interferometer using the 2.34 meter Vainu Bappu 
Telescope (VBT), situated at the Vainu Bappu Observatory (VBO), Kavalur, India.
\vspace{0.5cm}

\vspace{0.5cm}

\vspace{0.5cm}

\noindent
{\bf PLAN OF THE ARTICLE}
\vspace{0.3cm}

\noindent
{\bf 1. Prologue}

\noindent
{\bf 2. Interference of Two Light Waves}

2.1 The Case of Two Monochromatic Waves

2.2 The Case of Quasi-monochromatic Waves

\noindent
{3. A few Experiments to Measure Stellar Diameter}
	
\noindent
{4. Effect of Atmospheric Turbulence}
	
\noindent
{5. Speckle}
	
5.1 Imaging in the Presence of Atmosphere

5.2 Laboratory Simulation

5.3 Speckle Interferometer

5.4 Aperture Synthesis

5.5 Detector

\noindent
{6. Data Processing}
	
\noindent
{7. Image Processing}

7.1 Speckle Holography Method

7.2 Knox-Thomson Technique (KT)

7.3 Speckle Masking or Triple Correlation Technique (TC)

7.4 Relationships

7.5 Blind Iterative Deconvolution Technique (BID)

\noindent
{8. Astrophysical Programmes}

\noindent
{9. Epilogue}
\vspace{0.5cm}

\noindent
{\bf 1. Prologue}
\vspace{0.3cm}

\noindent
When a person looks at the medium size optical telescope of 2 meter class, he 
may wonder about its large structure rather than the complexity 
involved in making a good mirror. Much more complexity is in store for an
astronomer when he quests about the difficulties in obtaining
diffraction limited images. Where is the hindrance? Is it that the 
aberrations persist in the telescope or the atmospheric turbulence creating
the problem or both?
\vspace{0.3cm}

\noindent
Will they achieve the goal if a large telescope (diameter $>$ 2 meter) is 
installed? Large telescope helps in gathering more optical energy, as well as
in obtaining better angular resolution. The resolution increases with the
diameter of the aperture. Owing to the diffraction phenomenon, the image of
the point source (unresolved stars) cannot be smaller than a limit 
at the focal plane of the 
telescope. This phenomenon can be observed in ocean, when regular waves pass 
through an aperture. It is present in the sound waves, as well as in the
electro-magnetic spectrum too starting from gamma rays to radio waves.
\vspace{0.3cm}

\noindent
Though interferometry at optical wavelengths in astronomy began more than a 
century and a quarter ago [1], but real progress 
has been made at radio wavelengths in post war era. Development of long 
baseline and very long baseline interferometry (VLBI), as well as usage of 
sophisticated image processing techniques have brought high dynamic range
images with milliarcseconds (marcsec) resolution. 
\vspace{0.3cm}

\noindent
The understanding of the effect of atmospheric turbulence on the structure of 
stellar images and of ways to overcome this degradation has opened a channel 
to diffraction limited observations with large telescope in the optical domain.
Speckle interferometric technique [2] is being used to decode the
high angular resolution information. Diffraction limited information can also
be obtained with a large telescope using other techniques viz., (i) pupil plane 
interferometry [3], (ii) differential speckle 
interferometry [4], (iii) phase closure technique [5],
(iv) aperture synthesis using partial redundant, as well as 
non-redundant masking method [6 - 15], (v) speckle
spectroscopy [16], (vi) speckle polarimetry [17] etc.
\vspace{0.3cm}

\noindent
In the optical band, a large mirror of $\phi$ $>$10 meter class with a high 
precision accuracy in figuring may not be possible to develop, therefore, 
the resolution is restricted with the size of the telescope. Introduction
of long base line interferometry using diluted apertures became necessary
and the improvements of instrument technology made it possible to obtain a very 
high resolution of the order of a few marcsec [18, 19]. 
Recent success in implementing phase-closure technique on three ground 
based telescopes in the optical domain [20]
has produced new results with very high angular resolution.
\vspace{0.3cm}

\noindent
Developments of high resolution imaging have been going on in our institute 
over a decade. These were in the form of concentrating on theories of image
analysis, as well as on conducting several experiments at the various telescopes
and at the laboratory [21-26] and  Recently, we have developed a basic
speckle interferometer [27, 28] for use at the 2.34 meter
VBT, VBO, Kavalur.
\vspace{0.3cm}

\noindent 
Due to the paucity of funds, limitation 
of infrastructure facilities and the lack of support that includes man power 
etc., this project could not progress at a faster pace. Quite a bit of time 
had been spent on R \& D during the execution of the project [21, 22, 24, 27,
29-31]. A few vital 
optical components had to be developed by us [27]. Several 
algorithms were developed simultaneously with bare facilities to carry out 
image analysis and processing [23, 26, 32, 33]. 
Nevertheless, I could complete the mammoth task successfully. [I remember 
a few words of a scientist, 'one has to have an infinite patience to undertake 
any new challenging project and bring it to a success'. I do realize now 
the meaning of those words of caution which include stress and strain one has 
to undergo, that may lead to a loss of career (academic!). I may repeat the 
statement of Labeyrie [34], 'Recommendation to students: beware of large 
scale innovative projects, they can drive you crazy'].
\vspace{0.3cm}

\noindent
In the following, I shall talk here in brief, the basic 
principle of interferometry and its applications, several experiments conducted 
by the pioneers, as well as by us, the problems created by the atmosphere and 
the remedy, speckle interferometric technique and its use at the telescope in 
the optical domain, the salient features of our newly designed speckle 
interferometer and the detectors are being used for recording speckles etc. 
The application of non-redundant masking technique and its use 
at the telescopes are also discussed. This part of the talk is the concise 
version of the lecture
series which I delivered to the senior Ph. D. students at the fall of 
1995 at IIA, Bangalore. In addition, I shall discuss the various techniques 
applied to image restoration and their shortcomings, as well as the 
astrophysical problems which can be observed using 2.34 meter VBT, at Kavalur 
using the afore-mentioned interferometer. 
\vspace{0.5cm}

\noindent
2. {\bf Interference of Two Light Waves}
\vspace{0.3cm}

\noindent
When two light beams from a single source are superposed, the intensity at
the point of superposition varies from point to point between maxima which
exceed the sum of the intensities in the beams and minima, which may be zero, 
known as interference [35]. In the following, we shall discuss 
the degree of correlation that exists between the fluctuations in two light 
waves. 
\vspace{0.5cm}

\noindent
2.1 The Case of Two Monochromatic Waves:
\vspace{0.3cm}

\noindent
The intensity I of light has been defined as the time average of the amount
of energy which crosses in unit time, a unit area perpendicular to the 
direction of the energy flow. The electric vector {\bf E} is represented by

$${\bf E}_{(\bf r, t)} = R \left\{ {\bf A}_{({\bf r})} e^{-i \omega t} \right\}$$ 

\begin{equation}
= {\frac {1} {2}} \left[ {\bf A}_{({\bf r})} e^{-i \omega t} 
+ {\bf A}^{*}_{({\bf r})} e^{i \omega t} \right] 
\end{equation}

\noindent
here, ${\bf r}_{(x,y,z)}$ is a position vector of a point and ${\bf A}$ is a 
complex vector $a_j (\bf r) e^{ig_j (\bf r)}$. 
\vspace{0.3cm}

\noindent
For a homogeneous plane wave, the amplitudes $a_j's$ are constant, while in 
phase function $g_j$ are of the form

\begin{equation}
g_j (\bf r) = {\bf K}.{\bf r} - \delta_j 
\end{equation}

\noindent
where, ${\bf K}$ is the propagation vector and $\delta_j's$ are the phase 
constant, which specifies the state of polarization.

$${\bf E^2} = \left[ \frac{1} {2} \left\{ {\bf A}_{(\bf r)} e^{-i \omega t} + 
{\bf A}^{*}_{(\bf r)} e^{i \omega t} \right\} \right]^2$$  

\begin{equation}
= \frac {1} {4} \left( {\bf A}^2 e^{-2 i \omega t} + 
{\bf A}^{*2} e^{2i \omega t} + 2 {\bf A A}^{*} \right) 
\end{equation}

\noindent
whence, taking the time average over an interval large compared with the
period $T = 2\pi/\omega$
\vspace{0.3cm}

\noindent 
The intensity I,

$$I = <{\bf E^2}>$$ 

$$= \frac{1} {2} {\bf A.A}^{*}$$ 

$$= \frac{1} {2} \left( \mid A_x \mid^2 + \mid A_y \mid^2 + \mid A_z \mid^2 
\right)$$

\begin{equation}
= \frac{1} {2} \left( a_1^2 + a^2_2 + a^2_1 \right) 
\end{equation}

\noindent 
where, $< \ >$ is the ensemble average and $\ast$ stands for the complex 
conjugate.
\vspace{0.3cm}

\noindent 
Let us suppose now that two monochromatic waves {\bf E$_1$} and {\bf E$_2$} are
superposed at a point. The  total electric field at that point is

\begin{equation}
{\bf E} = {\bf E_1} + {\bf E_2} 
\end{equation}

\noindent
so that,

\begin{equation}
{\bf E^2} = {\bf E^2_1} + {\bf E^2_2} + 2{\bf E_1}.{\bf E_2} 
\end{equation}

\noindent 
The intensity at the same point is

\begin{equation}
I = I_1 + I_2 + J_{12} 
\end{equation}

\noindent
where $$I_1 = < {\bf E}^2_1 > \ , \ I_2 = < {\bf E}_2^2 >$$ are the intensities of
two waves and $$J_{12} = 2 < {\bf E_1} . {\bf E_2} >$$ is the interference term.
\vspace{0.3cm}

\noindent 
Let {\bf A} and {\bf B} be the complex amplitudes of the two waves, where 

\begin{equation}
A_x = a_1 e^{ig_1} \ ; \ B_x = b_1 e^{ih_1} 
\end{equation}

\noindent 
The (real) phases $g_j$ and  $h_j$ of the two waves will be different in general, 
since the two waves will have to travel to the intersecting point by different 
paths. If the same phase difference $\delta$ is introduced between the 
corresponding components, we have

\begin{equation}
g_1 - h_1 = g_2 - h_2 = g_3 - h_3 = \delta = \frac {2 \pi} {\lambda_o} 
\Delta \varphi 
\end{equation}

\noindent
$\Delta \varphi$ is the optical path difference (OPD) between two waves from
the common source to the intersecting point and $\lambda_o$ is the wavelength 
in vacuum.
\vspace{0.3cm}

\noindent 
In terms of {\bf A} and {\bf B}

$${\bf E_1} . {\bf E_2} = \frac {1} {4} \left( {\bf A} e^{-i \omega t} + 
{\bf A^*} e^{i \omega t} \right) . \left( {\bf B} e^{-i \omega t} + {\bf B^*} 
e^{i \omega t} \right)$$

\begin{equation}
= \frac{1} {4} \left( {\bf A} . {\bf B} e^{-2i \omega t} +
{\bf A^*} . {\bf B^*} e^{2i \omega t} + {\bf A} . {\bf B^*} + {\bf A^*} . 
{\bf B} \right) 
\end{equation}

\noindent 
Therefore,

$$J_{12} = 2 < {\bf E_1} . {\bf E_2} >$$ 

$$= \frac {1} {2} \left( {\bf A} . {\bf B^*} + {\bf A^*} . {\bf B} \right)$$

\begin{equation}
= \left( a_1 b_1 + a_2 b_2 + a_3 b_3 \right) \cos \delta 
\end{equation}

\noindent
This expression shows the dependence of the interference term on the amplitude
components and on the phase difference of the two waves. 
\vspace{0.3cm}

\noindent 
If the distribution of intensity resulting from the superposition of two waves
propagates in z$-$direction and linearly polarized with their {\bf E} vectors 
in the x$-$direction, then,

$$a_2 = a_3 = b_2 = b_3 = 0$$

\begin{equation}
I_1 = \frac {1} {2} a^2_1 \ ; \ I_2 = \frac {1} {2} b^2_1  
\end{equation}

\noindent 
and 

$$J_{12} = a_1 b_1 \ cos \delta$$ 

\begin{equation}
= 2 \sqrt{I_1 I_2} \ cos \delta 
\end{equation}

\noindent 
The total intensity

\begin{equation}
I = I_1 + I_2 + 2 \sqrt{I_1 I_2} \ cos \delta 
\end{equation}

\begin{equation}
I_{max} = I_1 + I_2 + 2 \sqrt{I_1 I_2} \;\;\;\;\;\;\;\  when  
\mid \delta  \mid \ = 0, 2 \pi, 4 \pi 
\end{equation}

\begin{equation}
I_{min} = I_1 + I_2 - 2 \sqrt{I_1 I_2} \;\;\;\;\;\;\;\ when  
\mid \delta \mid \ = \pi, 3 \pi, 5 \pi 
\end{equation}

\noindent 
when $I_1 = I_2$

\begin{equation}
I = 4 I_1 cos^2 \frac {\delta} {2}  
\end{equation}

\noindent 
The intensity varies between 4I and 0.
\vspace{0.5cm}

\noindent 
2.2 The Case of Quasi-monochromatic Waves:
\vspace{0.3cm}

\noindent 
In 1868, Fizeau found the relationship between the aspect of interference 
fringes and the angular size of the light source. The complex degree of 
co-herence $\gamma_{12} (\tau)$ of the observed source is defined as follows:

\begin{equation}
\gamma_{12} (\tau) = \frac {\Gamma_{12}(\tau)} {[\Gamma_{11} (0).
\Gamma_{22} (0)]^{1/2}} 
\end{equation}

\noindent
in which $$\Gamma_{12} (\tau) = < \psi_1 (t). \psi^\ast_2 (t + \tau)>$$ is the 
inter-correlation function of the field $\psi_1$ and $\psi_2$, measured at 
two points $r_1$ and $r_2$. 
\vspace{0.3cm}

\noindent
The ensemble average can be replaced by a time average due to the 
assumed ergodicity of the fields. $\Gamma_{11}(0)$ is the average intensity at 
point $r_1$.
\vspace{0.3cm}

\noindent
For quasi-monochromatic sources, the Van Cittert-Zernicke theorem states that in 
quasi-monochromatic light the modulus of the complex degree of coherence of the 
source is equal to the modulus of the normalized spatial Fourier transform of 
the source's brightness [35]. 

$$\mid \gamma_{12} (0) \mid = \frac {\mid  FT \ of \ brightness \ distribution 
\mid} {total \ intensity}$$

\begin{equation}
= \frac {\mid \widehat{O} (f_{12}) \mid} {\mid \widehat{O} (0) \mid} 
\end{equation}

\noindent
where, $| \ |$ stands for the modulus, $\widehat{ }$ \ is the Fourier 
transform and $\widehat{O}(f)$ is the FT of the object at spatial frequency f. 
\vspace{0.3cm}

\noindent
The modulus of the degree of coherence, sampled at several separations $(r_1, 
r_2)$ is the visibility function of the source which yields discrete values of 
the modulus of the energy spectrum of the source.
\vspace{0.3cm}

\noindent 
If both the fields are sent on a quadratic detector, it yields the desired 
cross-term (time average due to time response). The measured intensity at the 
detector would be 

$$I(r_1, r_2, \tau) = < \mid \psi_1 (t) + \psi_2 (t + \lambda) \mid^2 >$$

\begin{equation}
= I_1 + I_2 + 2 (I_1. I_2)^{1/2} .Re \{ \gamma_{12} (\tau) \} 
\end{equation}

\noindent
with $$I_1 = < \mid \psi_1 (t) \mid^2 >$$ and $$I_2 = <\mid \psi_2 (t) \mid^2 >$$
\vspace{0.3cm}

\noindent 
In order to keep the time correlation close to unity, the delay $\tau$ must
be limited to a small fraction of the temporal width $\tau_c$.

\begin{equation}
\Delta\nu \cdot \tau_c = 1 
\end{equation}

\noindent
where $\Delta\nu$ is the spectral width and $\tau_c$ is the temporal width.  
\vspace{0.3cm}

\noindent 
A factor less than unity affects the degree of coherence. The corresponding 
limit for the OPD between two fields is the coherence length, defined by

\begin{equation}
l_c = c. \tau_c = (\lambda_o)^2 / \Delta \lambda   
\end{equation}

\noindent 
If $\tau << \tau_c$, we can write,

\begin{equation}
\gamma_{12} (\tau) = \gamma_{12} (0) e^{-2 \pi i \nu_{o} \tau} 
\end{equation}

\noindent
where, [$\nu_{o}$ is mean frequency]
\vspace{0.3cm}

\noindent 
Let $\Phi_{12}$ be the argument of $\gamma_{12} (\tau)$ we have, 

\begin{equation}
I_{r_1, r_2, \tau} = I_1 + I_2 + 2 (I_1, I_2)^{1/2} Re \left\{ \mid \gamma_{12} 
(0) \mid e^{i \Phi_{12}} e^{-2 \pi i \nu_{o} \tau} \right\}  
\end{equation}

\noindent 
The measured intensity at a distance x from the origin (point at zero OPD) on 
a screen at distance x from the aperture is

\begin{equation}
I_{12} (x) = I_1 + I_2 + 2 (I_1, I_2)^{1/2} \mid \gamma_{12} (0)
\mid cos \left[\frac {2 \pi d(x)} {\lambda} - \Phi_{12} \right] 
\end{equation}

\noindent
where $d(x) = B.x/(z,\lambda)$ is the OPD corresponding to x. B is the distance 
between the two apertures.
\vspace{0.3cm}

\noindent 
The modulus of the spatial coherence of collected fields at the aperture appears
through the contrast of the fringes which can be measured by the following 
equation. 

$$C = \frac {I_{max} - I_{min}} {I_{max} + I_{min}}$$ 

\begin{equation}
= \mid \gamma_{12} (0) \mid \frac{2(I_1I_2)^{1/2}} {I_1 + I_2} 
\end{equation}

\noindent
where C is the visibility.
\vspace{0.3cm}

\noindent
In order to get $\mid \gamma_{12} (0) \mid$, the measurements of $I_1$ and $I_2$ 
should be made separately.
\vspace{0.5cm}

\noindent
3. {\bf A Few Experiments to Measure Stellar Diameter}
\vspace{0.3cm}

\noindent
In order to produce Young's fringes at the focal plane of the telescope, Fizeau 
[1] had suggested to install a screen with two holes on top of the telescope.
According to him, these fringes remain visible in presence of seeing, therefore,
allow measurements of stellar diameters with diffraction limited resolution.
Stefan attempted with 1 meter telescope at Observatoire de Marseille and 
fringes appeared within the common Airy disk of the sub-apertures. But he could 
not notice any significant drop of fringe visibility. Since the maximum 
achievable resolution is limited by the diameter of the telescope, he concluded 
none of the observed stars approached 0.1 arcsec. in angular size. 
\vspace{0.3cm}

\noindent
About half a century later, Michelson, who had spent some years with Fizeau,
could measure the diameter of the satellites of Jupiter with Fizeau 
interferometer on top of the Yerkes refractor. Similar interferometer (Fizeau 
mask) was placed on top of 100 inch telescope at Mt. Wilson [36] 
and the angular separation of spectroscopic binary star Capella was measured. We 
too conducted the same experiment at VBO, Kavalur using 1 meter Carl-Zeiss 
telescope and successfully recorded fringes of several bright stars with a 16 
mm movie camera giving an exposure of 16 msec. per frame [21].  
\vspace{0.3cm}

\noindent
To overcome the restrictions of the baseline Michelson [37] constructed his 
stellar interferometer by installing a 7 meter steel beam on top of the 
telescope. It was equipped with 4 flat mirrors to fold the beams in periscopic 
fashion. The supergiant star $\alpha-Orionis$ were resolved with this 
interferometer [38]. 
\vspace{0.3cm}

\noindent 
In this design, the maximum resolution is limited by the length of the girder
bearing the collectors. The spatial modulation frequency in the focal plane
is independent of the distance between the collectors. This feature allows to 
keep the same detection conditions when varying the baseline B. The telescope
serves as correlator, thus, provides zero OPD.
\vspace{0.3cm}

\noindent 
This experiment had faced various difficulties in resolving stars. These are 
mainly due to the (i) effect of atmospheric turbulence, (ii) variations of 
refractive index above small sub-apertures of the interferometer causing the
interference pattern to move as a whole, (iii) 7 meter separation of outer 
mirrors is insufficient to measure the diameter of more stars and 
(iv) mechanical instability prevents controlling large interferometer.
\vspace{0.5cm}

\noindent 
4. {\bf Effect of Atmospheric Turbulence}
\vspace{0.3cm}

\noindent
When an idealized astrophysical source of monochromatic radiation enters 
in the absence of atmosphere, is known as plane wave having uniform 
magnitude and phase across the telescope aperture. The point spread function
(PSF) of the telescope is the modulus square of the Fourier transform of the 
aperture function. The resolution at the image plane of the telescope is
determined by the width of the PSF. 
\vspace{0.3cm}

\noindent
When a flat wavefront passes down through atmosphere, it
suffers a phase fluctuations and reaches the entrance pupil of the telescope
with patches of random excursions in phase [39]. Due to the motion
and temperature fluctuations in the air above the telescope aperture,
inhomogeneities in the refractive index develop. These inhomogeneities have
the effect of breaking the aperture into cells with different values of
refractive index that are moved by the wind across the telescope aperture.
Kolmogrov law represents the distribution of turbule sizes, from millimeters 
to meters, with lifetimes varying from msecs to seconds. Changes in the 
refractive index in different portions of the aperture result to the
phase changes in the value of the aperture function. The time evolution of
the aperture function implies that the PSF is time dependent. 
If the atmosphere is frozen at a particular instant;
each patch of the wavefront with diameter $r_{o}$
$-$ Fried parameter $-$ would act independently of the rest of the wavefront 
resulting in many bright spots. These spots are known as speckles and spread
over the area defined by the long exposure image.
Computer simulated analysis demonstrates the destructions of the finer 
details of an image of a star by the atmospheric turbulence [23].
The size of $r_o$ is found to be varied between 8 to 12 cm at H$_\alpha$
wavelength during the night at the 2.34 meter VBT [40]. 
\vspace{0.3cm}

\noindent
The long-exposure PSF is defined by the ensemble average, $<S({\bf x})>$,
independent of any direction. The average illumination, $I({\bf x})$ of a
resolved object, $O({\bf x})$ obeys convolution relationship,

\begin{equation}
<I({\bf x})> = O({\bf x}) * <S({\bf x})>
\end{equation}

\noindent
where, * stands for convolution, ${\bf x}$ = (x,y) is a 2-dimensional space 
vector. Using 2-dimensional Fourier transform, this equation can be read as,

\begin{equation}
<\widehat{I}({\bf u})> = \widehat{O}({\bf u})\cdot<\widehat{S}({\bf u})>
\end{equation}

\noindent
where, $\widehat{O}({\bf u})$ is the object spectrum, $<\widehat{S}({\bf u})>$
is the transfer function for long-exposure images and is the product of the
transfer function of the atmosphere $B({\bf u})$, as well as the transfer 
function of the telescope, $T({\bf u})$. ${\bf u}$ is the
spatial frequency vector with magnitude u. The transfer function for 
long-exposure image can be expressed as,

\begin{equation}
<\widehat{S}({\bf u})> = B({\bf u})\cdot T({\bf u})
\end{equation}

\noindent
The benefit of the short-exposure images over long-exposure can be 
visualized by the following explanation.
Let us consider two seeing cells separated by a vector in the telescope pupil, 
$\lambda {\bf u}$, where $\lambda$ is the mean wavelength and ${\bf u}$ is an 
angular spatial frequency vector. If a point source is imaged through the 
telescope by using pupil function consisting of two apertures, corresponding
to the two seeing cells, then a fringe pattern is produced with narrow spatial 
frequency bandwidth. If the major component $I({\bf u})$ at the frequency 
${\bf u}$ is produced by contributions from all pairs of points with 
separations $\lambda {\bf u}$, with one point in each aperture and is averaged 
over many frames, then the result for frequencies greater than $r_o/\lambda$ 
tends to zero. The Fourier component performs a random walk in the complex
plane and average to zero, $<I({\bf u})>$ = 0, when $u > r_o/\lambda$.
\vspace{0.3cm}

\noindent
For a large telescope, the aperture, P, can be sub-divided into a set of
sub-apertures, $p_i$. According to the diffraction theory [35],
the image at the focal plane of the telescope is obtained by adding all such 
fringe patterns produced by all possible pairs of sub-apertures. With increasing
distance of the baseline between two sub-apertures, the fringes move with an 
increasingly larger amplitude. On a long-exposure images, no such shift is
observed, which implies the loss of high frequency components of the image.
While, in the short-exposure images ($<$20msec), the interference fringes are 
preserved. 
\vspace{0.5cm}

\noindent
5. {\bf Speckle}
\vspace{0.3cm}

\noindent
The term 'Speckle' refers to a grainy structure observed when an uneven surface
of an object is illuminated by a fairly coherent source. A good example of
speckle phenomena may be observed at the river port when many boats are
approaching towards the former at a particular time or in the swimming pool
when many swimmers are present. Each boat or swimmer emits wave trains and
interference between these random trains causes a speckled wave field on the
water surface. Depending on the randomness of the source, spatial or
temporal, speckles tend to appear. Spatial speckles may be observed when all
parts of the source vibrate at same constant frequency but with different
amplitude and phase, while temporal speckles are produced if all parts of it
have uniform amplitude and phase. With a non-monochromatic 
vibration spectrum, in the case of random sources of light, spatio-temporal 
speckles are produced.
\vspace{0.3cm}

\noindent
The ground illumination produced by any star has fluctuating speckles, known
as star speckles. It is too fast and faint, therefore, cannot be seen directly.
Atmospheric speckles can be observed easily in a star image at the focus of a 
large telescope using a strong eyepiece. The star image looks like a pan of
boiling water. If a short exposure image is taken, speckles can be recorded.
The speckle size is of the same order of magnitude as the Airy disc of
the telescope in the absence of turbulence. The number of correlation cells
is determined by the equation $N = D/r_{o}$. As the seeing improves, the 
number decreases.  
\vspace{0.8cm}

\noindent
5.1 Imaging in the Presence of Atmosphere :
\vspace{0.3cm}

\noindent
Let us consider an imaging system consists of a simple lens based telescope
in which the point spread function (PSF) is invariant to spatial shifts. An
object (point source) at a point ${\bf x}^\prime$ anywhere in the field of
view will, therefore, produce a pattern $S({\bf x} - {\bf x}^\prime)$ across
the image. If the object can emit incoherently, the image $I({\bf x})$ of a 
resolved object $O({\bf x})$ obeys convolution relationships. The 
mathematical description of the convolution of two functions is of the form:

\begin{equation}
I({\bf x}) = \int O({\bf x})^\prime S({\bf x} - {\bf x}^\prime) d{\bf x}^\prime
\end{equation}

\noindent
Convolution equations can be reduced to agreeable form using the Fourier 
convolution theorem. The Fourier transform of a convolution of two functions
is the product of the Fourier transform of the two functions. Therefore, in
the Fourier plane the effect becomes a multiplication, point by point, of
the transform of the object $\widehat{O}({\bf u})$ with the transfer function
$\widehat{S}({\bf u})$.
\vspace{0.5cm}

\noindent
5.2 Laboratory Simulation :
\vspace{0.3cm}

\noindent
Atmospheric seeing can be simulated at the laboratory by introducing disturbances
in the form of a glass plate with silicone oil [2]. We had 
introduced various static dielectric cells (SDC) of various sizes
etched in glass plate with hydrofluoric acid. Several glass plates with both
regular and random distribution of SDCs of known sizes were made and 
used in the experiment [31]. The phase-differences due to etching lie
between 0.2$\lambda$ and 0.7$\lambda$. In order to obtain
the light beam from a point source, similar to the star in the sky, we had 
developed an artificial star image by placing a pair of condensing lenses
along with micron-sized pin-hole in front of the source [22].
The beam was collimated with a good quality Nikkon lens; the wave fronts
from this artificial star enter a simulated telescope whose focal ratio is
1:3.25 (similar to the prime focus of VBT). The image was magnified to discern
the individual speckles with a high power microscope objective. 
The speckles were recorded through a 10nm interference
filter centered on 5577$\AA$. Figure 1 depicts (a) the laboratory set up to
simulate speckles from an artificial star, (b) speckles obtained in the 
laboratory through the aforementioned narrow band filter. The image was 
digitized with the PDS 1010M micro-densitometer and processed using the
COMTAL image processing system of the VAX 11/780 at the VBO, Kavalur. The 
clipping technique was used to enhance the contrast in grey levels. The clipped
image is superposed on the histogram-equalized original image. The laboratory
set up was sensitive enough to detect
aberrations produced by the objective lenses, as well as micro-fluctuations in
the speckle pattern caused by vibrations. By introducing an aperture mask in
front of this telescope, we could obtain the fringes [22, 29, 30]. 
The similarity of the observed image shape at the laboratory with 
the computer simulations was found. 
\vspace{0.3cm}

\noindent
\begin{figure}
\centerline{\psfig{figure=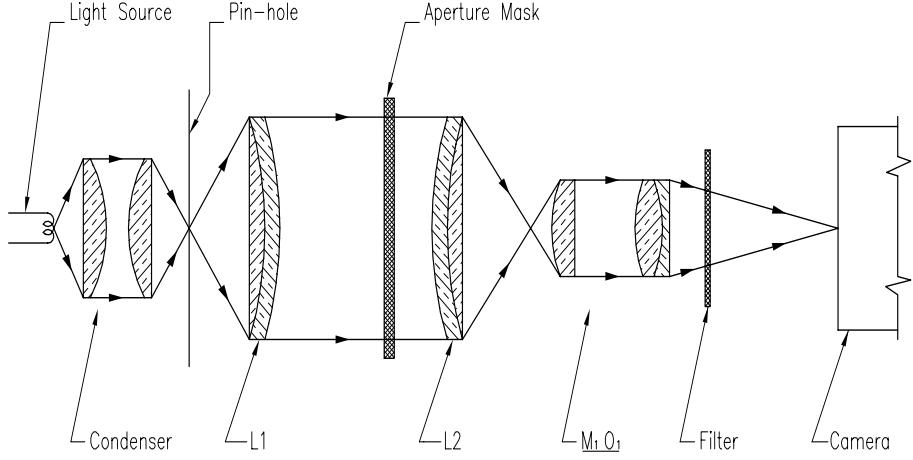,height=6cm,width=12cm}}
\caption{ (a). Laboratory set up to simulate speckles from an artificial star,} 
\end{figure}

\noindent
\begin{figure}
\centerline{\psfig{figure=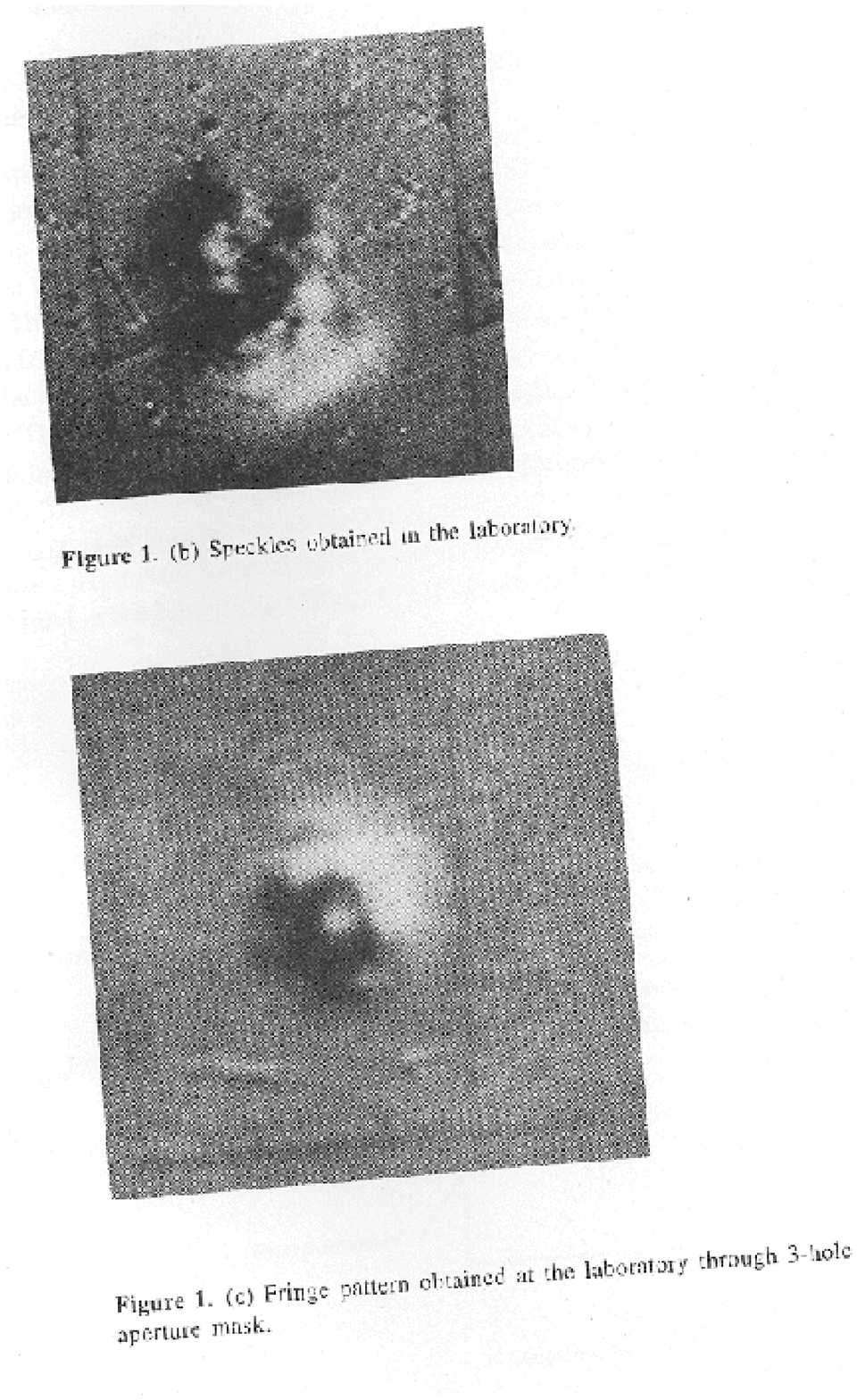,height=14cm,width=10cm}}
%\caption{ } 
\end{figure}

The laboratory simulation is necessary for the accurate evaluation of the
performance of speckle imaging system; comparison of the experimental results
should be made with the computer simulations. The importance of the systematic
use of simulated image is to validate the image processing algorithms in 
retrieving the diffraction limited information.
\vspace{0.5cm}

\noindent
5.3 Speckle Interferometer :
\vspace{0.3cm}

\noindent
A speckle interferometer is a high quality diffraction limited camera
where magnified ($\sim$ f/100) short exposure images can be recorded.
Additional element for atmospheric dispersion corrections is necessary
to be incorporated. At an increasing zenith distance speckles get elongated 
owing to this effect. Either a pair of Risley prism must be provided for
the corrections or the observation may be carried out using a narrow
bandwidth filter. In the following, the salient features of our newly developed 
speckle interferometer [27, 28] are described in brief. 
\vspace{0.3cm}

\noindent
The wave front falls on the focal plane of an optical flat made of low expansion 
glass with a high precision hole of aperture ($\sim$350 $\mu$), at an angle of 
15$^{o}$ on its surface [27]. The image of the object passes on 
to the microscope objective through this aperture, which slows down the image 
scale of this telescope to f/130. A narrow band filter would be placed before 
the detector, to avoid the chromatic blurring. The surrounding star field of 
diameter 10 mm, gets reflected from the optical flat on to a plane 
mirror and is re-imaged on an intensified CCD, henceforth ICCD [24].
We have recorded a large numbers of specklegrams of several
close binary systems and of other point source using uncooled ICCD as sensor. 
The image at the Cassegrain focus of 2.34 meter VBT is sampled to 0.015
arcsec per pixel. Figure 2 depicts (a) the optical layout of the interferometer,
(b) speckles of the close binary
star HR4689 obtained with the same set up at Cassegrain focus of the said 
telescope on 28$^{th}$-1$^{st}$ March, 1997. 
\vspace{0.3cm}

\noindent
\begin{figure}
\centerline{\psfig{figure=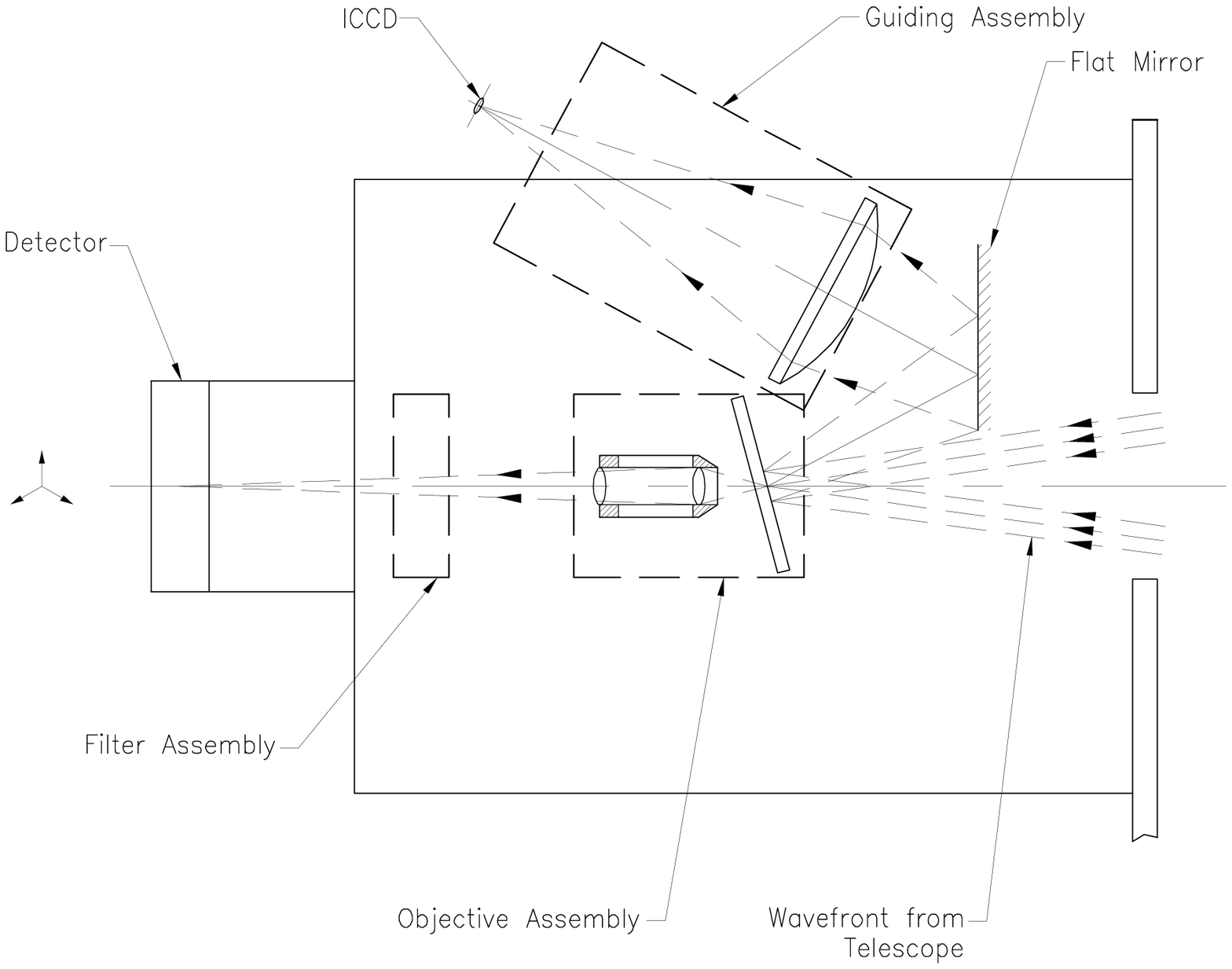,height=7cm,width=11cm}}
%\caption{(a)} 
\end{figure}

\noindent
\begin{figure}
\centerline{\psfig{figure=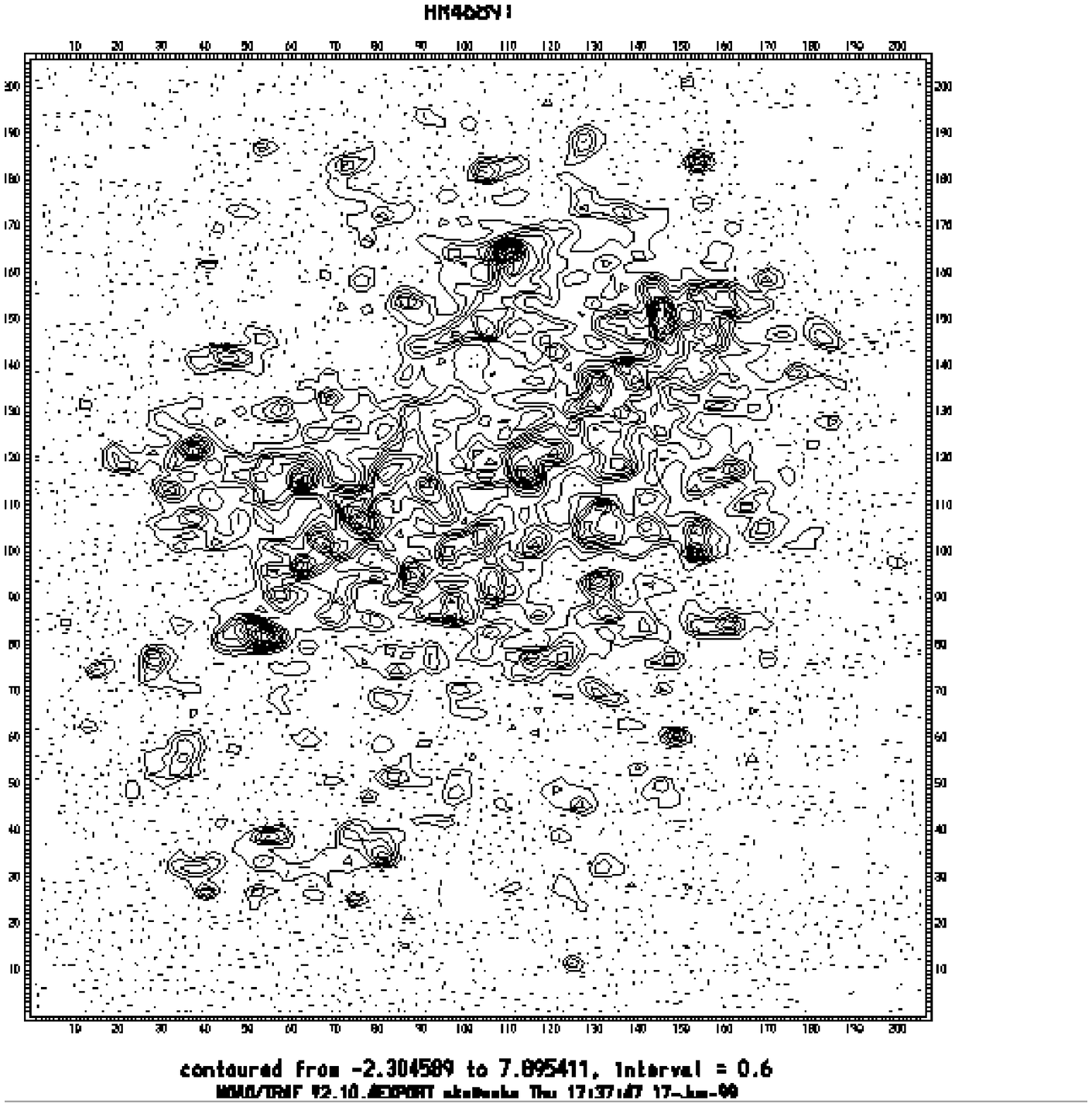,height=12cm,width=12cm,angle=270}}
\caption{ (a). Optical layout of the speckle interferometer,
(b). Speckles of the close binary star HR4689 obtained with 
the new speckle interferometer at VBT. The numbers on the axes denote pixel 
numbers with each pixel being equal to 0.015 arcsec. } 
\end{figure}

\noindent
\begin{figure}
\centerline{\psfig{figure=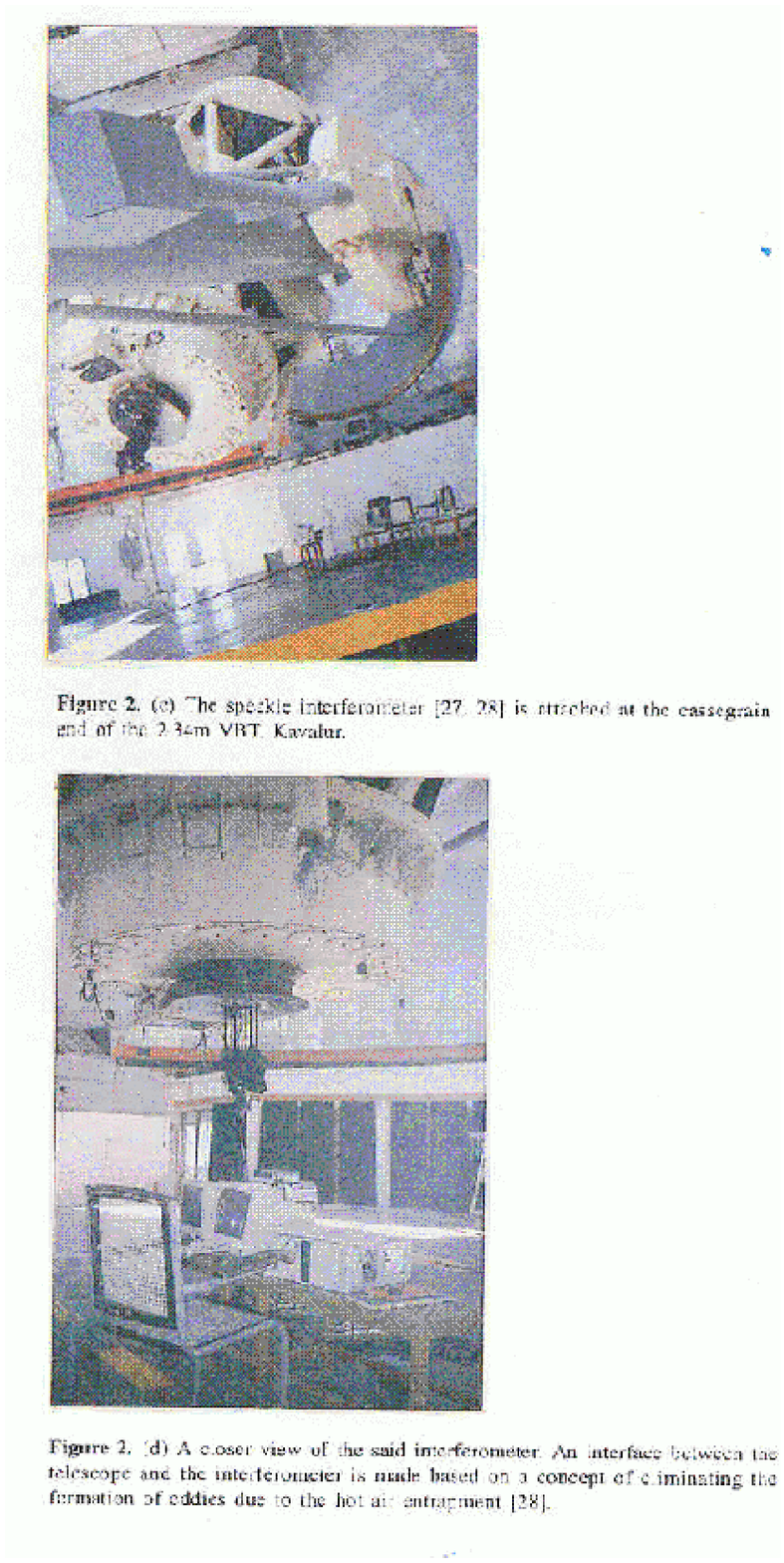,height=18cm,width=10cm}}
%\caption{ } 
\end{figure}

This interferometer is built with extreme care so to avoid
flexure problems which might affect high precision measurements of close binary
star systems etc., in an unfavourable manner. The design analysis has been 
carried out with the modern finite element method [41] and
computer aided machines were used in manufacturing to get dimensional and 
geometrical accuracies. The method requires the structure to be subdivided into
a number of basic elements like beams, quadrilateral and solid prismatic
elements etc. A complete structure is built up by the connection of such finite
elements to one another at a definite number of boundary points called
nodes and then inputting appropriate boundary constraints, material properties
and external forces. The relationship between the required deformations of the
structure and the known external forces is $[K]\{{\bf d}\} = {\{\bf F}\}$,
where, $[K]$ is the stiffness matrix of the structure, $\{{\bf d}\}$ is the
unknown displacement vector and $\{{\bf F}\}$ is the known force vector. All
the geometry and topology of the structure, material properties and boundary
conditions go into computation of $[K]$. The single reason for universal 
application of finite element method is the ease with which the matrix, $[K]$
is formulated for any given structure.
\vspace{0.5cm}

\noindent
5.4 Aperture Synthesis :
\vspace{0.3cm}

\noindent
Two promising methods, viz., (a) speckle masking (see chapter 7.3) technique 
[42], (b) non-redundant aperture masking technique [6, 7] 
are based on the principle of phase-closure 
method. To recover the Fourier phases of the source brightness distribution from 
the observations, it is necessary to detect fringes on a large baselines,
therefore, enables one to reconstruct images.
The concept of using three antennae arranged in a triangle was first
introduced in radio astronomy [43] in late fifties. Closure-phases
are insensitive to the atmospherically induced random phase errors, as well as
to the permanent phase errors introduced by the telescope aberrations in
optics. Since any linear phase term in the object cancels out [44], 
this method is insensitive to the position of the object but sensitive to 
any object phase non-linearity. 
\vspace{0.3cm}

\noindent
The measurements of the closure-phases was first obtained at high light level 
with three-hole aperture mask placed in the pupil plane of the telescope [5].
Interference patterns of the star were recorded using CCD as sensor. We had 
conducted similar experiment around the
same time by placing an aperture mask of 3-holes, 10cm in diameter, arranged in 
a triangle, over 1 meter telescope at VBO, Kavalur and tried to record the
interference pattern with our earlier version of the interferometer [21].
In this experiment, we used a 16mm movie camera as detector. 
The advantage of placing aperture mask over
the telescope, in lieu of pupil mask is to avoid additional optics. But
a curious modulation of intensity in the fringe pattern was noticed, therefore,
unable to proceed further. The modulation could result from a time-independent
aberration in the optical system of the telescope [21]. However,
we performed the similar experiment in the laboratory, using the aperture
mask of 3-holes, placed in front of a simulated telescope and were able to
record the interference patterns [30]. 
\vspace{0.3cm}

\noindent
The aperture synthesis imaging technique with telescope involves observing an
object through a masked aperture of several holes and recording the interference
patterns in a series of short-exposure. The patterns contain information about
structure of the object at the spatial frequencies from which an image of the
same can be reconstructed by measuring the visibility amplitudes and closure
phases. This method produces images of high dynamic range, but restricts to
bright objects. 
\vspace{0.3cm}

\noindent
Several groups have obtained the fringe patterns 
using both non-redundant and partially redundant aperture mask of N-holes
at large or moderate telescope [8-15]. In the 
laboratory, the shapes of fringe pattern of N-hole apertures were also studied 
by us by introducing the various aperture masks arranged in both non-redundant 
and partially redundantly [22, 29]. A few instruments developed for the pupil
plane aperture mask are described below.
\vspace{0.3cm}

\noindent
(i) The instrument developed for the Hale 5 meter telescope [11] 
used f/2.8 ($\phi$ 85mm) Nikkon camera lens to collimate f/3.3 primary
beam of the telescope. The lens formed an image of the primary mirror at 
a distance of about 85 mm where a mask was placed on a stepper-motor-driven
rotary stage controlled by a PC. Another identical lens forms a second focus
(scale is 12 arcsec/mm). This image was expanded by microscope objective 
(X80), enabling to sample 0.15 arcsec/mm on the detector. A narrow band 
interference filter (6300$\AA$, FWHM 30$\AA$) was placed between the microscope
objective and the detector. The resistive anode position sensing photon
counting detector [45] was used to record the interference 
patterns. They were successful in producing optical aperture synthesis maps of 
two binary stars $\beta-Corona Borealis$ and $\sigma-Hercules$.  
\vspace{0.3cm}

\noindent
(ii) University of Sydney [12] had developed a masked 
aperture-plane interference telescope (MAPPIT) for the 3.9 meter Anglo-Australian 
telescope to investigate interferometry with non-redundant masks.
A field lens re-images the telescope pupil down to diameter of 25 mm 
and the aperture mask is placed where the pupil image is formed. Dove prism is
used to rotate the field, allowing coverage of all position angles on the sky.
Dispersed fringes are produced using a combination of image
and pupil plane imaging. The camera lens and microscope objective produce
an image of a star in one direction. In the orthogonal direction, the 
detector receives the dispersed pupil image. The mask holes play
the role of the spectrograph slit. A cylindrical lens is used as the
spectrograph's camera lens. Image photon counting system is used in this 
experiment to record the dispersed fringe pattern. This instrument is used at
coud\'e focus of 3.9 meter Anglo-Australian telescope. They were able to
resolve several close binaries, as well as to measure angular diameters of
cool stars [13, 14].
\vspace{0.3cm}

\noindent
Recently, Bedding [15] of the said University has developed another version of 
this technique, called multiplexed one-dimensional speckle [MODS], by replacing 
holes in the aperture mask with a slits. 
Using a cylindrical lens that creates a continuous series of one-dimensional 
interferograms, interferograms from many arrays can be recorded side-by-side 
on a 2-dimensional detector. A narrow band filter is used in place of
dispersing prism. The mask containing slit
is placed in the collimating beam. The optics in the interference direction
form a image plane interferometer and in the orthogonal direction, the 
cylindrical lens produces a pupil image. Measurements at different position
angles can be made by rotating this lens. Two-dimensional image reconstruction
can be made from the series of one-dimensional interferograms [46].
Since the same number of photons will be collected in this technique as 
the conventional speckle interferometer, the limiting magnitude would not
change [15].
\vspace{0.8cm}

\noindent
5.5 Detector :
\vspace{0.3cm}

\noindent
The speckle interferometer requires snap shots of very high time resolution of 
the order of (a) frame integration of 50 Hz [47], (b) photon recording 
rates of a few MHz [48]. 
We have been using an uncooled ICCD as sensor to record  speckle-grams of 
the objects. It gives video signal as output. The images 
were recorded at an exposure 20 msec using a frame grabber card DT-2861
manufactured by Data Translation$^{TM}$. Each frame consists of odd and even
field. But the coherence 
time of the atmosphere is a highly variable parameter [49], it is 
desirable to observe speckles with a photon counting system. In the following, 
we shall describe the salient features of precision analogue photon address 
(PAPA) detector.
\vspace{0.3cm}

\noindent
The PAPA is a 2-d photon counting camera [48]. Individual 
photons hitting the photocathode of a high gain image intensifier 
produce a spot of light from a phosphor screen on the output side 
of the intensifier. An image of the phosphor screen is sent 
through optics to 19 photo-multiplier tubes (PMT), 18 of which have 
their active area covered by one of 9 different grey scale masks. The 19th
tube acts as an event strobe, registering a digital pulse if the spot
on the phosphor is detected by the instrument. 9 tubes are used to obtain
positional information for an event in one direction, while the other 9
are used for positional information in the orthogonal direction. If the 
phosphor hit in a region where it is not covered by the mask, an event 
is registered by the phototube. By taking the information from each 
phototube for each event, these are recorded as a list of photon
addresses and arrival times in a binary format. 
\vspace{0.5cm}

\noindent
6. {\bf Data Processing}
\vspace{0.3cm}

\noindent
Stellar speckle interferometry consists of taking many short integration images of
an object. The frame image is the convolution of the instantaneous PSF of
telescope$-$atmosphere $S(\bf x)$ with the actual object irradiance 
distribution $O(\bf x)$. The intensity distribution $I(\bf x)$ of the speckle 
interferograms, in the case of quasi-monochromatic incoherent source can be 
described by the following space-invariant imaging equations.

\begin{equation}
{I(\bf x) = {O(\bf x) \ast S(\bf x)}} 
\end{equation}

\noindent
In the Fourier domain, convolution becomes an ordinary product so that,

\begin{equation}
{\widehat{I}(\bf u) = {\widehat{O}(\bf u) \cdot \widehat{S}(\bf u)}} 
\end{equation}

\noindent
The ensemble averaged power spectrum is given by,

\begin{equation}
{<\mid \widehat{I}(\bf u) \mid^{2}> = {\mid \widehat{O}(\bf u) \mid^{2} 
\cdot <\mid \widehat{S}(\bf u) \mid^{2}>}} 
\end{equation}

\noindent
Since $\mid \widehat{S}(\bf u) \mid^{2}$ is a random function in which the 
detail is continuously changing, the ensemble average of this term becomes 
smoother. The smooth function can be performed on a point source yields $<\mid 
\widehat{S}(\bf u) \mid^{2}>$. The object autocorrelation can be obtained by 
inverse Fourier transform. Figure 3 depicts the auto-correlation of a binary 
star HR5747. The specklegrams of this star were
obtained at the Cassegrain focus of the 2.34m VBT using speckle interferometer
[21] on 16/17th March 1990. The uncooled ICCD was used as sensor
to record the specklegrams; each pixel of this ICCD was sampled to 0.026 arcsec.
\vspace{0.5cm}

\noindent
\begin{figure}
\centerline{\psfig{figure=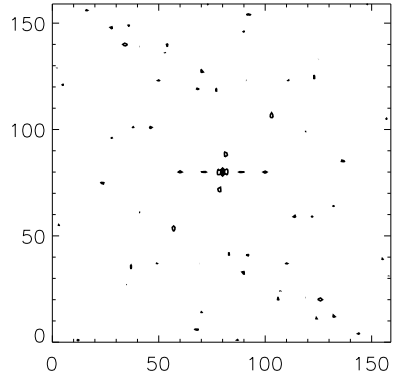,height=9cm,width=9cm}}
\caption{ Auto-correlation of a binary star HR5747. The speckles were recorded 
with earlier version of the speckle interferometer [21, 24]
on 16/17th March 1990. The numbers on the axes denote pixel numbers with each 
pixel being equal to 0.026 arcsec.}
\end{figure}

\noindent
7. {\bf Image Processing}
\vspace{0.3cm}

\noindent
Several methods have been developed to produce a map of the phase portion of the 
diffraction limited object Fourier transform. In the following we shall discuss 
some of the techniques.  
\vspace{0.5cm}

\noindent
7.1 Speckle Holography :
\vspace{0.3cm}

\noindent
I. In speckle holographic technique [50] complete image 
reconstruction becomes possible when a reference print source is available 
in the field of view (within isoplanatic patch $<$ 7 arcsec).
\vspace{0.3cm}

\noindent
Let us represent the point source by a Dirac impulse $A \delta(\bf x)$ at the 
origin and let $O_1(\bf x)$ be a nearby object to be reconstructed. The 
irradiance distribution in the field of view is 

\begin{equation}
O(\bf x) = A \delta (\bf x) + O_1 (\bf x)
\end{equation}

\noindent
A regular speckle interferometric measurement will give the squared modulus of 
its Fourier transform $\widehat{O}({\bf u})$

$$\mid \widehat{O} (\bf u) \mid^2 = \mid A + \widehat{O}_1 (\bf u) \mid^2$$

\begin{equation}
= A^2 + A \widehat{O}_1 (\bf u) + A \widehat{O}_1^* (\bf u) + 
\widehat{O}_1 (\bf u) \widehat{O}_1^* (\bf u) 
\end{equation}

\noindent 
The inverse Fourier transform gives the autocorrelation $C_o (\bf x)$ 
of the field of view

\begin{equation}
C_o {(\bf x)} = A^2 \delta {(\bf x)} + A O_1 {(\bf + x)} + A O_1 {(\bf - x)}
+ C_{o1} {(\bf x)} 
\end{equation}

\noindent
where $C_{o1}{(\bf x)}$ is the autocorrelation of the object.  
\vspace{0.3cm}

\noindent
The first and the last term in equation (36) are centered at the origin. If the 
object is far enough from the reference source, $O(\bf x)$ its mirror image 
$O{(\bf - x)}$ is therefore, recovered apart from a 180$^\circ$ rotation 
ambiguity. 
\vspace{0.3cm}

\noindent
II. The alternate idea is to calculate the average cross spectrum between the
objects and the reference. Let $O_1{(\bf x)}$ and $O_2{(\bf x)}$ be respectively the object and
reference brightness distributions and $I_1 {(\bf x)}$ and $I_2{(\bf x)}$ their 
associated instantaneous image intensity distributions. The relation between 
the objects and the images in the Fourier space becomes

\begin{equation}
\widehat{I}_1 (\bf u) = \widehat{O}_1 (\bf u) \cdot \widehat{S} (\bf u) 
\end{equation}

\begin{equation}
\widehat{I}_2 (\bf u) = \widehat{O}_2 (\bf u) \cdot \widehat{S} (\bf u) 
\end{equation}

\noindent
where, $\widehat{S} {(\bf u)}$ is the impulse response of the telescope
and the atmosphere. 
\vspace{0.3cm}

\noindent
The average cross-spectrum between the object and the reference 
\vspace{0.3cm}

\begin{equation}
< \widehat{I}_1 {(\bf u)} \widehat{I}_2^* {(\bf u)} > = \widehat{O}_1 {(\bf u)}
\widehat{O}_2^* {(\bf u)} \cdot < \mid \widehat{S} {(\bf u)} \mid^2 >_M 
\end{equation}

\noindent
The equation (39) shows that as the speckle holography transfer function 
$< \mid \widehat{S} {(\bf u)} \mid^2 >$ is real, the method is insensitive to 
aberrations and the phase of the cross spectrum expected value coincides with
the phase difference between the object and the reference.
\vspace{0.5cm}

\noindent
7.2 Knox-Thomson Technique (KT) :
\vspace{0.3cm}

\noindent
The method [51] involves in interfering $I({\bf u})$ with
itself after translating by a small shift vector ${\bf \Delta u}$.
The KT correlation may be defined in Fourier space as products of $\widehat{I}
({\bf u})$.

\begin{equation}
\widehat{I}({\bf u}, {\bf \Delta u}) = <\widehat{I}{(\bf u)} \widehat{I}^* 
({\bf u} + {\bf \Delta u})> 
\end{equation}

\noindent
This gives us the product,

\begin{equation}
<\widehat{I}{(\bf u)} \widehat{I}^* ({\bf u} + {\bf \Delta u})> 
= \widehat{O}({\bf u})\widehat{O}^{*}({\bf u}+{\bf \Delta u}) \cdot <\widehat{S}
({\bf u})\widehat{S}^{*}({\bf u}+{\bf \Delta u})>
\end{equation}

\noindent
Introducing the phases explicitly in the form $ \widehat{O} = \mid 
\widehat{O}\mid e^{j \phi_{o}}$ etc. and using ${\bf \Delta \phi} = \phi 
({\bf u}) - \phi ({\bf u}_1+{\bf \Delta u})$, we have   

\begin{equation}
\mid\widehat{I}{(\bf u)}\mid \mid \widehat{I}({\bf u} + {\bf \Delta u})\mid e^{j
\phi_{I}} = \mid\widehat{O}{(\bf u)}\mid \mid \widehat{O}({\bf u} + {\bf \Delta 
u})\mid e^{j \phi_{O}} \cdot \mid\widehat{S}{(\bf u)}\mid \mid \widehat{S}({\bf 
u} + {\bf \Delta u})\mid e^{j \phi_{S}}
\end{equation}

\noindent
If this equation is averaged over a large number of frames, the feature $({\bf
\Delta \phi_{S}}) = 0$. When ${\bf \Delta u}$ is small, $\mid \widehat{O}({\bf 
u}+{\bf \Delta u}) \mid \ \approx \ \mid \widehat{O}({\bf u}) \mid$ etc. and so

\begin{equation}
<\widehat{I}{(\bf u)} \widehat{I}^* ({\bf u} + {\bf \Delta u})> = \mid\widehat
{O}({\bf u})\mid^{2} e^{j \Delta \phi_{O}} \cdot (a \ smooth \ function)
\end{equation}

\noindent
from which, together with equation (32), ${\bf \Delta \phi_{O}}$ can be 
determined.
\vspace{0.5cm}

\noindent
7.3 Speckle Masking or Triple Correlation Technique (TC) :
\vspace{0.3cm}
 
\noindent
In case of non-availability of reference point source within iso-planatic 
patch, the instantaneous PSF can be estimated from the speckle pattern itself.
Weigelt [52] suggested to multiply the object speckle pattern $I({\bf x})$ by 
an appropriately shifted version of this $I({\bf x} + {\bf x}_1)$. For a binary
star, the shift is equal to the angular separation between the stars, 
masking one of the two component of each double speckle. The result is 
correlated with $I({\bf x})$.    
The Fourier transform of the triple correlation is called bispectrum and its
ensemble average [42] is given by

\begin{equation}
\widehat{I}({\bf u}_1, {\bf u}_2) = <\widehat{I}({\bf u}_1) \widehat{I}^{*} 
({\bf u}_1 + {\bf u}_2) \widehat{I}({\bf u}_2)> 
\end{equation}

\noindent
where, $\widehat{I}({\bf u}_1)$, $\widehat{I}({\bf u}_2)$, $\widehat{I}({\bf u}_1 
+ {\bf u}_2)$, denote the Fourier transforms of $I({\bf x})$. 
\vspace{0.3cm}

\noindent
In the second order moment or in the energy spectrum, phase information is lost,
but in the third order moment or in the bispectrum it is preserved. If we put
equation (32) into equation (44), it emerges as,

\begin{equation}
\widehat{I}({\bf u}_1, {\bf u}_2) = \widehat{O}({\bf u}_1)\widehat{O}^{*}(
{\bf u}_1+{\bf u}_2)\widehat{O}({\bf u}_2) <\widehat{S}({\bf u}_1)\widehat{S}
^{*}({\bf u}_1+{\bf u}_2)\widehat{S}({\bf u}_2)>
\end{equation}

\noindent
The relationship shows that the image bispectrum is equal to the object 
bispectrum times a bispectral transfer function. The object bispectrum is
given by,

$$\widehat{I}_{o}({\bf u}_1, {\bf u}_2) = \widehat{O}({\bf u}_1) \widehat{O}
^{*} ({\bf u}_1 + {\bf u}_2) \widehat{O}({\bf u}_2) $$ 

\begin{equation}
= \frac{ < \widehat{I}({\bf u}_1) \widehat
{I}^{*} ({\bf u}_1 + {\bf u}_2) \widehat{I}({\bf u}_2) >} {< \widehat{S}({\bf 
u}_1) \widehat{S}^{*} ({\bf u}_1 + {\bf u}_2) \widehat{S}({\bf u}_2) >}  
\end{equation}

\noindent
The modulus $\mid \widehat{O}({\bf u}) \mid$ of the 
object Fourier transform $\widehat{O}({\bf u})$ can be derived from the object 
bispectrum $\widehat{I}_{o}({\bf u}_{1},{\bf u}_{2})$ [53].  
\vspace{0.3cm}

\noindent
The phase $\phi({\bf u})$ of the object Fourier transform can also be derived
from the object bispectrum. Let $\phi_b$ be the phase of the object bispectrum 
and we get,

\begin{equation}
\widehat{O}({\bf u}) = \mid \widehat{O}({\bf u}) \mid e^{j \phi ({\bf u})} 
\end{equation}

\noindent
and

\begin{equation}
\widehat{I}_{o}({\bf u}_1,{\bf u}_2) = \mid \widehat{I}_{o}({\bf u}_1,{\bf u}_2)
\mid e^{j \phi_b ({\bf u}_1,{\bf u}_2)} 
\end{equation}

\noindent
Equations (47) and (48) may be inserted into equation (42), therefore, yields
the relations,

\begin{equation}
\widehat{I}_{o}({\bf u}_1,{\bf u}_2) = \mid\widehat{O}({\bf u}_1)\mid e^{j \phi 
({\bf u}_1)}\mid\widehat{O}({\bf u}_2 )\mid e^{j \phi ({\bf u}_2)}\mid\widehat
{O}({\bf u}_1+{\bf u}_2)\mid e^{-j \phi ({\bf u}_1+{\bf u}_2)} \rightarrow  
\end{equation}

\begin{equation}
e^{j \phi_b ({\bf u}_1,{\bf u}_2)} = e^{j \phi ({\bf u}_1)} e^{j \phi ({\bf u}_2)}
e^{-j \phi ({\bf u}_1+{\bf u}_2)} \rightarrow  
\end{equation}

\begin{equation}
\phi_b({\bf u}_1,{\bf u}_2) = \phi ({\bf u}_1) + \phi ({\bf u}_2)
- \phi ({\bf u}_1+{\bf u}_2) \rightarrow  
\end{equation}

$$\phi ({\bf u}_1+{\bf u}_2) = \phi ({\bf u})$$ 

\begin{equation}
= \phi ({\bf u}_1) + \phi ({\bf u}_2) - \phi_b ({\bf u}_1, {\bf u}_2)  
\end{equation}

\noindent
Equation (52) is a recursive equation for calculating the phase of the object 
Fourier transform at coordinate ${\bf u} = {\bf u}_1 + {\bf u}_2$ [53].
If the object spectrum at ${\bf u}_1$ and ${\bf u}_2$ are known, the
phase spectrum at $({\bf u}_1 + {\bf u}_2)$ can be computed. But the bispectrum
phases are of mod $2\pi$, therefore, the reconstruction in equation (48) may lead
to $\pi$ phase mismatches between the computed phase-spectrum values along
the different paths to the same point in frequency space. The unit amplitude
phasor recursive re-constructor are insensitive to the phase ambiguities and
the computing argument of the term, $e^{j\phi({\bf u}_1+{\bf u}_2)}$, can be
expressed as,   

\begin{equation}
e^{j\phi({\bf u}_1 + {\bf u}_2)} = e^{j[\phi({\bf u}_1) + \phi({\bf u}_2) -
\phi_b ({\bf u}_1,{\bf u}_2)]} 
\end{equation}

The triple correlation algorithm based on this unit amplitude recursive 
re-constructor method was developed at our institute for processing the stellar 
objects [33], as well as for the extended object [54].
\vspace{0.5cm}

\noindent
7.4 Relationships :
\vspace{0.3cm}

\noindent
From the sections 7.2 and 7.3, we find the relationship among the two widely
used algorithms, namely, KT and TC methods [55].  
In autocorrelation technique, the major Fourier component of the fringe pattern 
is averaged as a product with its complex conjugate and so the atmospheric 
phase contribution is eliminated and the averaged signal is non-zero (see
chapter 4). Unfortunately the phase information is not preserved. 
\vspace{0.3cm}

\noindent
The KT is a small modification of the autocorrelation technique 
In the KT method, approximate phase closure is achieved by two vectors, 
${\bf u}$ and ${\bf u}+{\bf \Delta u}$, and assuming that the pupil phase is 
constant over ${\bf \Delta u}$. The major Fourier component of the fringe 
pattern is averaged with a component at a frequency displaced by a vector 
${\bf \Delta u}$. The vector displacement ${\bf \Delta u}$ should not force
the vector difference ${\bf -u -\Delta u}$ outside the spatial frequency
bandwidth of the fringe pattern which, in turn, preserves the Fourier phase
difference information in the averaged signal. The atmospheric phase 
effectively forms a closed loop.
\vspace{0.3cm}

\noindent
In the bispectrum method a third vector, ${\bf \Delta u}$ is added to form 
phase closure. When $\lambda \Delta u > r_o$, the third vector $\bf \Delta u$
is essential. The KT method fails with this arrangement. In this system, the
two apertures are extended to three and the $\Delta u > r_o/\lambda$. 
It can be seen that atmospheric phase contribution is not closed. KT is limited 
to frequency differences $\Delta u < r_o/\lambda$. If the bispectrum average
is performed, the phase is closed and Fourier phase difference information is
preserved. This method can obtain phase information for phase difference 
$\Delta u > r_o/\lambda$. 
\vspace{0.5cm}

\noindent
7.5 Blind Iterative Deconvolution Technique (BID) :
\vspace{0.3cm}

\noindent 
In this technique, the iterative loop is repeated enforcing image-domain and 
Fourier-domain constraints until two images are found that produce the input 
image when convolved together [56, 57]. 
The image-domain constraints of non-negativity is generally used in iterative 
algorithms associated with optical processing to find effective supports of the 
object and or PSF from a speckle-gram. The implementation of the BID, used by us 
is described below.
\vspace{0.3cm}

\noindent
The algorithm has the degraded image $I({\bf x})$ as the operand. An initial
estimate of the PSF, $S({\bf x})$, has to be provided. The degraded image is 
deconvolved from the guess PSF by Wiener filtering, which is an operation
of multiplying a suitable Wiener filter ( constructed from the Fourier
transform $\widehat{S}({\bf u})$ of the PSF) with the Fourier transform 
$\widehat{I}({\bf u})$ of the degraded image. The technique of Wiener filtering
damps the high frequencies and minimizes the mean square error between each 
estimate and the true spectrum. This filtered deconvolution takes the form

\begin{equation}
\widehat{O}({\bf u}) = \widehat{I}({\bf u)}{\frac{\widehat{O}_f({\bf u})}
{\widehat{S}({\bf u})}}
\end{equation}

\noindent
The Wiener filter, $\widehat{O}_f({\bf u})$, is given by the following
equation:

\begin{equation}
{\widehat{O}_f({\bf u}) = {\frac {\widehat{S}({\bf u})\widehat{S}^{*}({\bf u})}
{\mid\widehat{S}({\bf u})\mid^2 + \mid\widehat{N}({\bf u})\mid^2}}} 
\end{equation}

\noindent
This noise term, $\widehat{N}({\bf u})$, can easily be replaced with a constant
estimated as the rms fluctuation of the high frequency region in the spectrum,
where the object power is negligible. The Wiener filtering spectrum, 
$\widehat{O}({\bf u})$, takes the form:

\begin{equation}
{\widehat{O}({\bf u}) = \widehat{I}({\bf u}) {\frac {\widehat{S}^{*}({\bf u})}
{\widehat{S}({\bf u}) \widehat{S}^{*}({\bf u}) + \widehat{N}({\bf u})
\widehat{N}^{*}({\bf u})}}} 
\end{equation}

\noindent
This result $\widehat{O}(\bf u)$ is transformed to image space, the negatives 
in the image are set to zero, and the positives outside a prescribed domain 
(called object support) are set to zero. The average of negative intensities
within the support are subtracted from all pixels. The process is repeated 
until the negative intensities decreases below the noise.
\vspace{0.3cm}

\noindent
A new estimate of the PSF is next obtained by Wiener filtering the original
image $I({\bf x})$ with a filter constructed from the constrained object
$O({\bf x})$. This completes one iteration. This entire process is repeated
until the derived values of $O({\bf x})$ and $S({\bf x})$ converge to sensible 
solutions. Before applying this scheme of BID to the real data, we have 
tested the algorithm with computer simulated convolved functions of binary star
and the PSF caused by the atmosphere and the telescope. The reconstruction
of the Fourier phase of these are shown in figure 4. The starting guess for
the PSF used for this calculation was a Gaussian with random noise. We were able
to obtain the output image, as well as the output PSF after 225 iterations.
\vspace{0.3cm}

\noindent
\begin{figure}
\centerline{\psfig{figure=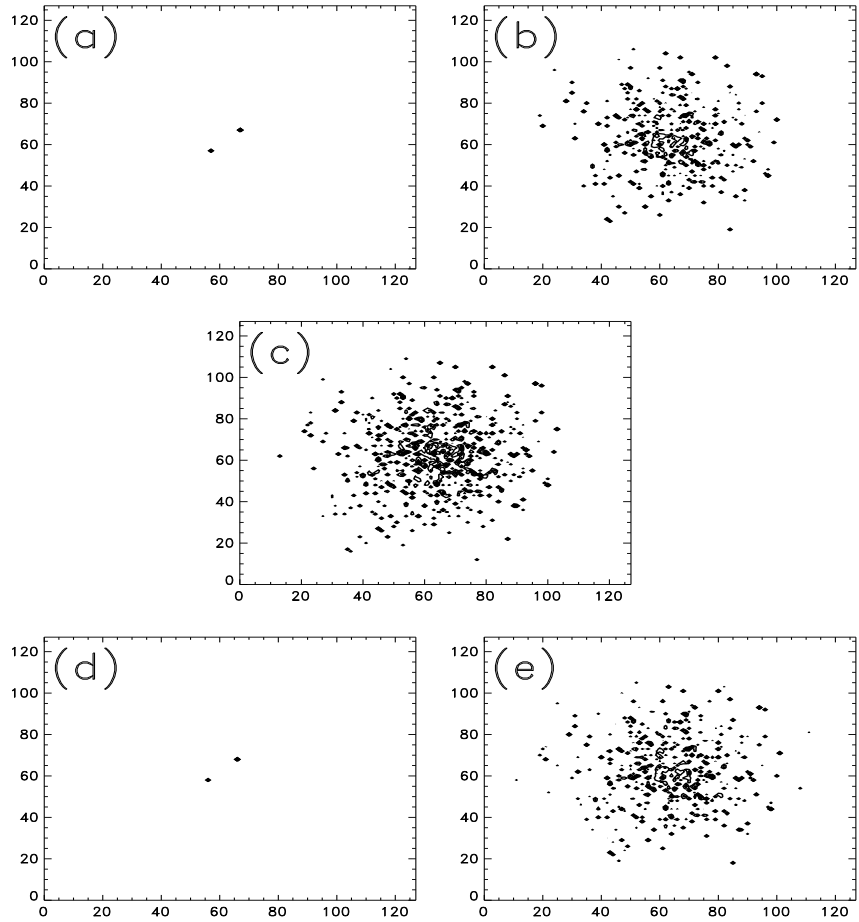,height=14cm,width=13cm}}
\caption{ 2-dimensional maps of the simulated image of (a) binary star, 
(b) simulated atmospheric PSF, (c) the convolved functions of these two (bpsf),
(d) the retrieved image of the binary star, (e) reconstructed atmospheric PSF.}
\end{figure}

\noindent
The uniqueness and convergence properties of the deconvolution algorithm
are uncertain for the evaluation of the reconstructed images if one uses
the BID method directly. We have tested the code [58] and found
that it is essential to estimate the input support radius of the object using
the auto-correlation technique, which helps in completing the reconstruction 
satisfactorily after several iterations [32].
\vspace{0.5cm}

\noindent
8. {\bf Astrophysical Programmes}
\vspace{0.3cm}

\noindent
The contribution of single aperture interferometry to several important fields 
in astrophysics has increased considerably, viz., separation and orientation of 
close binary stars [32, 59 - 61], shapes of 
asteroids [62], mapping of the finer features of extended objects 
[26], sizes of certain types of circumstellar envelopes [63, 64], 
structures of active galactic nuclei [65, 66],
resolving the gravitationally lensed 
QSO's [67] etc. We too plan to observe the following interesting objects, 
if an intensified photon counting detector [48] can be 
made available.  
\vspace{0.3cm}

\noindent
(i) Studies of binary stars play a fundamental role in measuring stellar masses,
providing the benchmark for stellar evolution calculations. A long term
benefit of speckle imaging is a better calibration of the main-sequence
mass-luminosity relationship. Most measurements have been made at large aperture 
telescopes by groups in France, Russia, and the United States. Programmes of 
binary star interferometry are being carried out at telescopes of moderate and
small aperture too. But measurements from the southern hemisphere continue to be
rare. Many rapidly moving southern binaries are being ignored and a number of 
discoveries are yet to be confirmed.
\vspace{0.3cm}

\noindent
(ii) Most of the late-type stars are available in the vicinity of sun. All 
known stars, within 5 pc radius from the sun are red dwarfs with $m_v >$ +15.
Due to the intrinsically faint nature of K- and M- dwarfs, their physical
properties are not studied extensively. These dwarfs may often be close
binaries which can be detected by speckle interferometric technique.
\vspace{0.3cm}

\noindent
(iii) Another important field of observational astronomy is the studies of the 
physical processes, viz., temperature, density and velocity of gas in the
active region of the active galactic nuclei (AGN). Optical imaging in the light 
of emission lines on sub-arcsec scales can reveal the structure of the 
narrow-line gas. The scale of narrow-line regions is well resolved by the 
diffraction limit of a moderate-sized telescope.
\vspace{0.3cm}

\noindent
(iv) The spatial distribution of circumstellar matter surrounding objects
which eject mass, particularly young compact planetary nebulae can be determined 
[63, 64]. 
\vspace{0.3cm}

\noindent
(v) Capability of resolving the gravitationally lensed QSO's in the range of
0.2 arcsec. to 0.6 arcsec. will allow the discovery of many more lensing
events [67].
\vspace{0.5cm}

\noindent
9. {\bf Epilogue}
\vspace{0.3cm}

\noindent
The understanding of the basic random interference phenomenon $-$ speckle $-$
is of paramount importance to the observational astronomy. In recent years the
uses of speckle pattern and a wide variety of applications have been found 
in many other branches of physics and engineering. Though the
statistical properties of speckle pattern is complicated, detailed analysis  
of this is useful in information processing. Though the stellar speckle
interferometry is capable of detecting relatively faint objects ($\sim$ 16th. 
magnitude), the angular resolution is limited by the diameter of the telescope.
Angular resolution can vastly be improved by using long base line
interferometry using two or more telescopes. 
\vspace{0.3cm}

\noindent
India has an outstanding group in the field of radio astronomy using
long baseline interferometry. From the experience we have gained in developing
the field of optical interferometry, we are confident in building a long 
baseline interferometer in the optical and IR band. Long baseline 
interferometric observations of the objects [34, 68]
would offer the possibilities for direct measurement of all the basic 
physical parameters for a large number of stars.  
\vspace{0.5cm}

\noindent
{\bf References}
\vspace{0.3cm}

\noindent
[1]. H. Fizeau, 1868, C. R. Acad. Sci. Paris, {\bf 66}, 934.

\noindent
[2]. A. Labeyrie, 1970, Astron. and Astrophys., {\bf 6}, 85.

\noindent
[3]. C. Roddier and F. Roddier, 1988, Proc. NATO-ASI, 'Diffraction Limited 
Imaging with Very Large Telescopes', ed. D M Alloin and J M Mariotti, 
Carg$\grave{e}$se, France, 221.

\noindent
[4]. R. Petrov, F. Roddier, and C. Aime, 1986, J. Opt. Soc. Am. A., {\bf 3}, 634.

\noindent
[5]. J. E. Baldwin, C. A. Haniff, C. D. Mackay, and P. J. Warner, 1986, Nature, 

\noindent
[6]. D. H. Rogstad, 1968, App. Opt. {\bf 7}, 585.

\noindent
[7]. W. T. Rhodes and J. W. Goodman, 1973, J. Opt. Soc. Am., {\bf 63}, 647.

\noindent
[8]. D. F. Busher, C. A. Haniff, J. E. Baldwin and P. J. Warner, 1990, 
Mon. Not. Roy. Astro. Soc., {\bf 245}, 7.

\noindent
[9]. C. A. Haniff, C. D. Mackay, D. J. Titterington, D. Sivia, J. E. Baldwin
and P. J. Warner, 1987, Nature, {\bf 328}, 694.

\noindent
[10]. C. A. Haniff, D. F. Busher, J. C. Christou and S. T. Ridgway, 1989,  
Mon. Not. Roy. Astro. Soc., {\bf 241}, 51.

\noindent
[11]. T. Nakajima, S. R. Kulkarni, P. W. Gorham, A. M. Ghez, G. Neugebauer, 
J. B. Oke, T. A. Prince and A. C. S. Readhead, 1989, A J., {\bf 101}, 1510.

\noindent
[12]. T. R. Bedding, J. G. Robertson, R. G. Marson, P. R. Gillingham, R. H. 
Frater and J. D. O'Sullivan, 1992, Proc. ESO-NOAO conf. 'High Resolution Imaging 
Interferometry', ed., J. M. Beckers \& F. Merkle, Garching bei M\"unchen, FRG,
391.

\noindent
[13]. T. R. Bedding, J. G. Robertson and R. G. Marson, 1994, Astron. Astrophys., 
{\bf 290}, 340.

\noindent
[14]. T. R. Bedding, A. A. Zijlstra, O. Von der L\"he, J. G. Robertson, 
R. G. Marson, J. R. Barton and B. S. Carter, 1997, 
Mon. Not. Roy. Astro. Soc., {\bf 286}, 957.

\noindent
[15]. T. R. Bedding, 1999, astro-ph/9901225, Pub. Astron. Soc. Pacific (to appear).

\noindent
[16]. F. Grieger and G. Weigelt, 1992, Proc. ESO-NOAO conf. 'High Resolution Imaging 
Interferometry', ed., J M Beckers and F Merkle, Garching bei M$\ddot{u}$nchen, 
FRG, 225.
 
\noindent
[17]. H. Falcke, K. Davidson, K. H. Hofmann and G. Weigelt, 1996, Astron. 
Astrophys., {\bf 306}, L17.

\noindent
[18]. A. Labeyrie, 1975, Astrophys. J., {\bf 196}, L71.

\noindent
[19]. A. Labeyrie, G. Schumacher, M. Dugu$\acute{e}$, C. Thom, P. Bourlon, F. Foy, 
D. Bonneau, and R. Foy, 1986, Astron. and Astrophys., {\bf 162}, 359.

\noindent
[20]. J. E. Baldwin, R. C. Boysen, C. A. Haniff, P. R. Lawson, C. D. Mackay, 
J. Rogers, D. St-Jacques, P. J. Warner, D. M. A. Wilson, J. S. Young, 1998, 
Proc. SPIE., on 'Astronomical Interferometry', {\bf 3350}, 736.
{\bf 320}, 595.

\noindent
[21]. S. K. Saha, P. Venkatakrishnan, A. P. Jayarajan and N. Jayavel, 1987,
Curr. Sci. {\bf 56}, 985.

\noindent
[22]. S. K. Saha, A. P. Jayarajan, K. E. Rangarajan and S. Chatterjee, 1988, Proc.
ESO-NOAO conf. 'High Resolution Imaging Interferometry', ed. F. Merkle,
Garching bei M$\ddot{u}$nchen, FRG, 661.

\noindent
[23]. P. Venkatakrishnan, S. K. Saha and R. K. Shevgaonkar, 1989, Proc., 'Image
Processing in Astronomy', ed. T. Velusamy, 57.

\noindent
[24]. V. Chinnappan, S. K. Saha and Faseehana, 1991, Kod. Obs. Bull. {\bf 11}, 87.

\noindent
[25]. S. K. Saha, B. S. Nagabhushana, A. V. Ananth and P. Venkatakrishnan, 
1997, Kod. Obs. Bull., {\bf 13}, 91. 

\noindent
[26]. S. K. Saha, R. Rajamohan, P. Vivekananda Rao, G. Som Sunder, R. Swaminathan
and Lokanadham, B., 1997, Bull. Astron. Soc. Ind., {\bf 25}, 563.

\noindent
[27]. S. K. Saha, A. P. Jayarajan, G. Sudheendra, and A. Umesh Chandra, 1997, 
Bull. Astron. Soc. Ind., {\bf 25}, 379.

\noindent
[28]. S. K. Saha, G. Sudheendra, A. Umesh Chandra, and V. Chinnappan, 1998, 
Experimental Astronomy (to appear).

\noindent
[29]. S. K. Saha, 1990, VBT News, No., 3, 3.

\noindent
[30]. S. K. Saha, 1991, IIA Newsletter, {\bf 6}, 11.

\noindent
[31]. S. K. Saha, 1991, VBT News, No., 8 \& 9, 9.

\noindent
[32]. S. K. Saha and P. Venkatakrishnan, 1997, Bull. Astron. Soc. Ind., {\bf 25}, 
329.

\noindent
[33]. S. K. Saha, R. Sridharan and K. Sankarasubramanian, 1999, 'Speckle image
reconstruction of Binary Stars', Presented at XIX ASI meeting held at Bangalore.

\noindent
[34]. A. Labeyrie, 1988, Proc. NATO-ASI, 'Diffraction Limited Imaging with Very Large
Telescopes', ed. D M Alloin and J M Mariotti, Carg$\grave{e}$se, France, 327.

\noindent
[35]. M. Born and E. Wolf, 1975, Principles of Optics, Pergamon Press.

\noindent
[36]. J. A. Anderson, 1920, Astrophys. J., {\bf 51}, 263.

\noindent
[37]. A. A. Michelson, 1920, Astrophys. J., {\bf 51}, 257.

\noindent
[38]. A. A. Michelson and F. G. Pease, 1921, Astrophys. J., {\bf 53}, 249.

\noindent
[39]. D. C. Fried, 1966, J. Opt. Soc. Am., {\bf 56}, 1972.

\noindent
[40]. S. K. Saha and V. Chinnappan, 1999, Bull. Astron. Soc. Ind. (to appear).

\noindent
[41]. O. C. Zienkiewicz, 1967, 'The Finite Element Methods in Structural and
Continuum Mechanics', McGrawhill Publication.

\noindent
[42]. A. W. Lohmann, G. P. Weigelt and A. Wirnitzer, 1983, Appl. Opt. {\bf 22}, 
4028.

\noindent
[43]. R. C. Jennison, 1958, Mon. Not. Roy. Astron. Soc., {\bf 118}, 276.

\noindent
[44]. S. K. Saha, 1999, 'Emerging Trends of Optical Interferometry in Astronomy',
communicated to Bull. Astron. Soc. India.

\noindent
[45]. M. Clampin, J. Croker, F. Paresce and M. Rafal, 1988, Rev. Sci. Instru. {\bf 59},
1269.

\noindent
[46]. D. F. Busher and C. A. Haniff, 1993, J. Opt. Soc. Am A., {\bf 10}, 1882.
 
\noindent
[47]. A. Blazit, 1986, Proc., 'Image Detection and Quality' - SFO, ed., SPIE,
{\bf702}, 259.

\noindent
[48]. C. Papaliolios, P. Nisenson and S. Ebstein, 1985, App. Opt. {\bf 24}, 287.

\noindent
[49]. R. Foy, 1988, Proc., 'Instrumentation for Ground Based Optical Astronomy - 
Present and Future', ed., L. Robinson, Springer Verlag, New York, 345.

\noindent
[50]. Y. C. Liu, and A. W. Lohmann, 1973, Opt. Comm., {\bf 8}, 372.

\noindent
[51]. K. T. Knox, and B. J. Thompson, 1974, Astrophys. J. {\bf 193}, L45.

\noindent
[52]. G. Weigelt, 1977, Opt. Communication, {\bf 21}, 55.

\noindent
[53]. G. Weigelt, 1988, Proc. NATO-ASI, 'Diffraction Limited Imaging with Very Large
Telescopes', ed. D. M. Alloin and J. M. Mariotti, Carg$\grave{e}$se, France, 191.

\noindent
[54]. R. Sridharan and P. Venkatakrishnan, 1999, Presented at XIX ASI meeting held at 
Bangalore.

\noindent
[55]. G. R. Ayers, M. J. Northcott and J. C. Dainty, 1988, J. Opt. Soc. Am. A, 
{\bf 5}, 963.

\noindent
[56]. G. R. Ayers and J. C. Dainty, 1988, Opt. lett., {\bf 13}, 457.

\noindent
[57]. R. H. T. Bates and B. L. K. Davey, 1988, Proc. NATO-ASI, 'Diffraction Limited 
Imaging with Very Large Telescopes', ed. D. M. Alloin and J. M. Mariotti, 
Carg$\grave{e}$se, France, 293.

\noindent
[58]. P. Nisenson, 1992, Proc. ESO-NOAO conf. 'High Resolution Imaging Interferometry', 
ed., J M Beckers and F Merkle, Garching bei M$\ddot{u}$nchen, FRG, 299.
 
\noindent
[59]. H. A. McAlister, 1988, Proc. ESO-NOAO conf. 'High Resolution Imaging 
Interferometry', ed. F. Merkle, Garching bei M$\ddot{u}$nchen, FRG, 3. 

\noindent
[60]. H. A. McAlister, W. I. Hartkopf, B. D. Mason and M. M. Shara, 1996, Astron. J, 
{\bf 112}, 1169.

\noindent
[61]. W. I., Hartkopf, B. D. Mason, H. A. McAlister, N. H. Turner, B. I. Barry, 
O. G. Franz and C. M. Prieto, 1996, Astron. J, {\bf 111}, 936.
   
\noindent
[62]. J. Drummond, A. Eckart and E. K. Hege, 1988, Icarus, {\bf 73}, 1.

\noindent
[63]. M. J. Barlow, B. L. Morgan, C. Standley, and H. Vine, 1986, Mon. 
Not. Roy. Astron. Soc., {\bf 223}, 151.

\noindent
[64]. P. R. Wood, S. J. Meatheringham, M. A. Dopita and D. M. Morgan, 1987, 
Astrophys. J., {\bf 320}, 178.

\noindent
[65]. V. S. Afanas'jev, I. I. Balega, Y. Y. Balega, V. A. Vasyuk and V. G. Orlov,
1988, Proc. ESO-NOAO conf. 'High Resolution Imaging Interferometry', 
ed., F Merkle, Garching bei M$\ddot{u}$nchen, FRG, 127. 

\noindent
[66]. S. Ebstein, N. P. Carleton and C. Papaliolios, 1989, Astrophys. J., {\bf 336}, 103.

\noindent
[67]. R. Foy, 1992, Proc. ESO-NOAO conf. 'High Resolution Imaging Interferometry', 
ed., J M Beckers and F Merkle, Garching bei M$\ddot{u}$nchen, FRG, 5. 

\noindent
[68]. D. Mourard, I. Bosc, A. Labeyrie, L. Koechlin, and S. Saha, 1989, Nature,
{\bf 342}, 520.
\vspace{0.5cm}

\noindent
{\bf ABOUT THE REVIEWER}
\vspace{0.3cm}

\noindent
Dr. Swapan K. Saha had received his Ph. D (Tech.) from the Institute of 
Radiophysics and Electronics, Calcutta University in 1983 and is a scientist at 
the Indian Institute of Astrophysics, Bangalore, India. In mid eighties, his 
interest had focused on to the 'Optical interferometry in astronomy' and spent 
a year at Observatoire de la Cote d'Azur (formerly C E R G A), Caussols, France, 
to study the high angular features of stars, viz., resolving emission envelope 
of $\gamma-Cassiopae$, eclipsing binary Algol A - B system etc.,  
using GI2T (Grand Interf\'erom\`etre \`a deux t\'elescope), an optical 
interferometer with a pair of 1.5 metre telescopes on a North-South baseline. 
He has a strong background in experimental Physics 
and developed various equipments for carrying out research in different arenas. 
Among others, the speckle interferometer for the 2.34 meter Vainu Bappu 
Telescope, Vainu Bappu Observatory, Kavalur, an exceptional one,
is used regularly to study the high resolution features of 
different types of celestial objects, viz., close binary stars, active galactic 
nuclei, Proto-planetary nebulae etc.; image reconstruction algorithms which
preserve the phase in the object Fourier transform, namely, triple correlation 
technique, blind iterative deconvolution technique are applied to map the 
atmospherically degraded objects. He is the author of several papers and 
is a member of International Astronomical Union, as well as a member of 
Astronomical Society of India. 
\end{document}